\documentclass[11pt]{article}
\usepackage{setspace}
\usepackage{amssymb}
\usepackage{amsmath}
\usepackage{amsfonts}

\def\be{\begin{equation}}

\def\ee{\end{equation}}

\newcommand\vecbf[1]{{\bf #1}}

\def\dd{\partial}

\def\bea{\begin{eqnarray}}

\def\eea{\end{eqnarray}}

\makeatletter
\def\blfootnote{\xdef\@thefnmark{}\@footnotetext}
\makeatother

\begin{document}

\singlespace

\begin{flushright} BRX TH-6305 \\
CALT-TH 2016-006
\end{flushright}

\vspace*{.3in}

\begin{center}

{\Large\bf Massive to gauge field reduction and gravitational wave zone information}

{\large S.\ Deser}

{\it 
Walter Burke Institute for Theoretical Physics, \\
California Institute of Technology, Pasadena, CA 91125; \\
Physics Department,  Brandeis University, Waltham, MA 02454 \\
{\tt deser@brandeis.edu}
}
\end{center}

\begin{abstract}
I analyze the possible relevance of LIGO's gravitational wave detection to the viability of massive gravity models. In GR, a wave zone, where the linearized approximation holds, is guaranteed to exist and the observed wave's amplitude profile can be sufficiently related to the emitting strong field interior to verify that, in this case, it was due to an inspiraling black hole merger. After an excursion to massive spin $1$'s massless limit, linear massive tensor theory is shown explicitly to propagate only (retarded) maximal, helicity $2$, modes to $O(m)$ as $m\rightarrow 0$; however, we don't know if the full theory has a similar ``wave zone" governed by the linear model. Even if it does, a much more serious obstacle for massive gravity is to construct a time-varying strong field event to compare with the strong field footprint of LIGO's observed signals. 
\end{abstract}

\section{Introduction}
LIGO's [1] recent observations, both verify (already at amplitude, rather than intensity, level) GR's gravitational radiation predictions and corroborate that its profile agrees in detail with calculated emission by an inspiraling black hole merger in the strong field emission region [2]. They even provide upper bounds on any putative linear massive model's mass using dispersion arguments. The aim of the present study is to discuss ways by which the data might bound the observational viability of these models, which are already afflicted by serious theoretical problems:  Even massive spin $1$ non-abelian, QCD, generalizations of massive (Proca) QED are known to be non-renormalizable [3], while massive gravities suffer, already classically, from local acausality -- the existence of local (unlike Godel spaces') closed timelike curves, amongst other ills [4], at non-linear, nonabelian, level.

A first step will be analysis of the massless limit of massive spin $1$ Proca theory, as 
a prelude to that of its Fierz-Pauli (FP) [5] spin $2$ counterpart. In both cases, I will show that only the maximal helicity modes stay coupled to their respective vector current and stress tensor sources to $O(m)$, and that these maximal helicity signals propagate according to the massive propagator $\Delta(x)$, that smoothly limits to the massless $D(x)$. However, the limit is not entirely smooth: indeed, the respective degree of freedom counts remain discontinuous, there being $2$, namely helicity $\pm s$ for all $m=0$ spin $s >0$, but $2s+1$ for their massive counterparts.
There is little to report on the interior, strong emitting field in the massive case:  it is even unknown whether a ``wave zone" exists there, unlike that proved to exist in GR [6]. Further, there are as yet no time-dependent strong field calculations with which to compare. This is likely to be a difficult enterprise: even the meaning of the Vainshtein mechanism [7] is rather obscure there [8]. Unless there is a true wave zone, and a successful accounting of the observed amplitude by some as yet unknown interior process, massive gravity will have been observationally rejected on the basis of LIGO's observation. 

\section{Proca limits}
Some six decades ago, Bass and Schrodinger [9] examined Abelian massive 
vector theory in an attempt to show that Maxwell electrodynamics is not isolated, but rather the smooth endpoint of a continuum of massive one. Their results were not quite complete (nor quite correct), so one part of this work is to provide them in full generality. However, this is for us just a useful introduction to the same question for general relativity's (GR) linearized spin $2$, tensor field: is it also a smooth endpoint of the massive Fierz-Pauli (FP) model [5]? We will indeed establish continuity in most aspects, but also find an unavoidable discontinuity in the number of field excitations, however small the mass. In addition, one must require, by fiat, that all sources be conserved, rather than by consistency conditions required of $m=0$, gauge field sources by Bianchi identities. An obvious symptom that non-conserved currents are problematic as $m \rightarrow 0$ is that 
\begin{equation*}
\dd^\mu \, A_\mu \sim m^{-2} \, \dd_\mu \, j^\mu \hbox{ and } \dd^\mu \, \left( h_{\mu\nu} - \eta_{\mu\nu} \hbox{tr} h \right) \sim m^{-2} \dd_\mu \, T^{\mu\nu}.
%?^µ A_µ ~m^-2  ?_µ j^µ and ?^µ (h_µnu -eta_µ nu tr h)~ m^-2 ?^µ T^µ nu. 
\end{equation*}

For spin one, this is may be a reasonable requirement on electric currents, but not for (even linearized) spin $2$, whose $T_{\mu\nu}$ is no longer conserved, precisely by virtue of its interaction with the field!  Of course we now know that the authors of [9] labored in an Abelian paradise and that both the Yang-Mills and massive nonlinear extensions of GR are profoundly discontinuous in the $m\rightarrow 0$ limit, as noted in the Introduction. Indeed, even Abelian spin $2$ is only continuous in an (A)dS context [10]: the famous vDVZ [11] discontinuity in the effective matter-matter interactions generated by FP can only be avoided by introducing another mass-dimension ``regulating" parameter, the cosmological constant, that can only be set to $0$ after taking the $m\rightarrow 0$ limit.
 We now show that the third, helicity $0$, massive vector degree of freedom essentially decouples (but does not disappear!) from its sources and ceases to radiate to $O(m)$ as the mass vanishes; to this end, we use the ADM ``$3+1$" and orthogonal $3$-space decompositions for spatial vectors and tensors a la [12]. Our signature is mostly plus. The Proca action is 
\begin{eqnarray}
I_m[A] &=& \int d^4 x \, \left[ -1/4 \, F_{\mu\nu}^2 - 1/2 \, m^2 \, A_\mu^2 + A_\mu \, j^\mu\right]  \\
&=& I_m[A^T_i] + I_m[A^L_i, A_0] \nonumber \\
&\equiv& 1/2 \, \int d^4 x \left[ ({\dot A}^T_i)^2 - (A^T_i)^2 \, (m^2 - \nabla^2) \, A^T_i + 2 \, A^T_i \, j^T_i \right]  \nonumber \\
&+& 1/2 \, \int d^4 x \left[ ({\dot A}^L_i)^2 - m^2 ((A^L_i)^2 - A_0^2) +( \nabla A_0)^2 - 2 \, \dot A^L_i \, \dd_i \, A_0 + 2 \, A^L_i \, j^L_i + 2 \, A_0 \, j^0 \right]  \nonumber
% \\
%T_{\mu\nu} &=& F_{\mu \alpha} \, F_\nu^{\, \, \, \alpha} - 1/4 \, \eta_{\mu\nu} \, F_{\alpha\beta}^2 + m^2 \, (A_\mu \, A_\nu - 1/2 \, \eta_{\mu\nu} \, A_\alpha^2) \\
%E_m &=& \int d^3 x \, T_{00} = 1/2 \, \int d^3 x \left[ (\dd_0 A_i - \dd_i  A_0)^2 + m^2 (A_i^2 + A_0^2)\right].
\end{eqnarray}
This being a linear system, it was easy to separate the helicity $1$ and $0$ components by decomposing the $3$-vector fields and sources. The latter are taken to be external and conserved:
\begin{eqnarray}
A_i &=& A^T_i + A^L_i, \, \, \, \, \, \, \, \, \, \, \, \nabla \cdot \vecbf A^T = 0 = \nabla \times \vecbf A^L, \, \, \, \, \, \, \, \, \, \, \,  \, \, \, \, \, \, \, \, \, \, \,
A_i^L \equiv \frac{\dd_i}{\sqrt{-\nabla^2} } \, a \\
\dd_\mu j^\mu &=& 0 \equiv \nabla \cdot \vecbf j + \dd_0 \, j^0 \equiv \sqrt{-\nabla^2} j + \dot\rho.
\end{eqnarray}
It also follows from their definitions that the inner products of any two different helicity fields, $V^T$ and $W^L$, are orthogonal under spatial integration because they are orthogonal projections of (vector) unity, as is obvious in Fourier space. In particular, this means that the total action consists of two separate parts, as given in (1).

The $A_i^T$, helicity $1$, action's only $m$-dependence is through the massive, rather than massless, propagator $(\Box-m^2)^{-1}$, so the limit indeed smoothly reduces to the Maxwell action in all its radiative aspects, namely dipole waves emitted by the transverse currents, and their retarded interactions mediated by the (slightly massive) transverse ``photon". [Indeed, LIGO obtained an upper bound on $m$ from dispersive analysis in the wave zone.] The Coulomb/Yukawa instantaneous interaction will emerge presently, as the (only) survivor of the helicity $0$ sector. 
The remaining, longitudinal action is, from (1),
\begin{equation}
\begin{aligned}
I[a] &= 1/2 \int d^4 x \left[ \dot a^2 - m^2 \, a^2 + 2 a j\right] - 1/2 \int d^4 x \left[ \rho \, Y \rho - 2 \, \rho Y \sqrt{-\nabla^2} \dot a + \dot a Y(-\nabla^2) a \right];  \\
Y &\equiv (m^2 - \nabla^2)^{-1},
\end{aligned}
\end{equation}
after completing squares in the $A_0^2 +A_0$ constraint sector of (1), to remove its $A_0$ dependence. It now remains to establish the consequences of this helicity zero action; this requires repeated use of current conservation, (3). We expect, following [9], the massless limit to be smooth and yield a free helicity $0$ excitation\footnote{Note in this connection that this otherwise inert mode remains unavoidably coupled to gravity, so that if had somehow been copiously produced at the Big Bang -- but, unlike photons, not by radiating charges -- it would forever exist as truly inert ``dark matter": since gravity couples to all matter, it will also do so with the helicity $0$ part of Proca, something omitted in [9].}, together with the standard instantaneous Coulomb interaction.

I merely sketch the steps: Redefining a (time-locally) in terms of a new variable S,
\begin{equation}
a = m^{-1}\, \sqrt{m^2 - \nabla^2} \, S
\end{equation}
eventually yields
\begin{equation}
I(S)= - 1/2 \int d^4x [S \, \Box \, S ]+ 1/2 \int d^4x \rho \nabla^{-2} \rho +O(m)  
\end{equation}
as expected. While the redefinition (5) is singular at $m=0$, this simply reflects the unavoidable discontinuity of degree of freedom count between Proca and Maxwell. Indeed, the only way to understand the amount of radiation of these longitudinal waves is to first put their action into normalized form (6), that is with a normal free field part. To emphasize this requirement in a familiar context, if we alter the normalization of the Maxwell kinetic term in QED to say $L_{\hbox{\tiny{Max}}} = - c^2/4 \, F_{\mu\nu}^2 +A_\mu \, j^\mu$, even with a finite constant $c$, this would amount to rescaling the current by $c^{-1}$ with respect to its normalized value. The net result then is that Proca indeed limits smoothly to Maxwell plus a free helicity $0$ mode that only couples to gravity.
The above results can also be read off from the effective current-current action among conserved currents,
\begin{equation}
I_m(j) = 1/2 \int d^4 x j^\mu \Delta \, j_\mu, \, \, \, \, \, \, \, \, \, \, \, \Delta \equiv (\Box - m^2)^{-1}
\end{equation}
The $j^T$ parts indeed couple with retardation (and of course $\Delta$ reduces smoothly to the massless propagator $D$), while the, ostensibly also retarded, remainder reduces straightforwardly, upon using conservation, precisely to (6): specifically the$\int  j^L \, \Delta \,  j^L$ terms combine with $\int \rho  \, \Delta \, \rho$ to form a numerator $ \sim  \int \rho\, Y\, \rho$, $Y=(\nabla^2 -m^2)^{-1}$.

\section{Massive spin 2}
  We now turn to our main interest -- spin $2$ -- and the $m \rightarrow 0$ relation of FP to linearized, $m=0$, GR. Rather than subject the reader to dissection of the full action, whose free part is already messy enough, and has already been separated into its helicity sectors in, e.g., [13], we concentrate on the source-source, $T_{\mu\nu}-T^{\mu\nu}$ effects, which is of course equivalent to reading off the propagator. The free, kinematical part consists of the five, helicity $(\pm 2,\pm 1,0)$ sectors, the $3$ lower ``longitudinal" ones again non-interacting in the $m=0$ limit, while the helicity $2$ propagates according to $\Delta$, i.e., with $D$ on the light cone in the limit. The $T-T$ action is trivial to obtain because the numerators of the FP (or gravity) propagator involve projectors whose $\dd_\mu \, \dd_\nu$ terms vanish when acting on the, conserved by fiat, $T_{\mu\nu}$. We first encounter here the ancient vDVZ [11] ``paradox", that the the respective numerators differ by a constant, even in the massless limit\footnote{An easy way to understand why there is a discontinuity is that the sum of 
the trace and double divergence of the massive field equations yields the discontinuous field-current identity [14] tr$ h_{\mu\nu} \sim  m^{-2} \hbox{tr} T_{\mu\nu}$.}:
\begin{equation}
I[m\rightarrow 0;m = 0]  = 1/2 \int d^4x [T_{\mu\nu} \, D \, T^{\mu\nu} - (1/3;1/2)\,  \hbox{tr} T \, D\,  \hbox{tr} T].   
\end{equation}
As is well-known, if the wrong, $1/3$, limit is used in (8), [11] showed that the bending of light would be off by $25\%$ from its GR value in order to keep the Newtonian value of $G$.

As mentioned above, it was later discovered [10] that the above discontinuity could be avoided by keeping a non-vanishing cosmological constant: the massive propagator's second term magically reverts to $1/2$  from $1/3$, while the massless one stays at $1/2$. [One can take the Minkowski, $\Lambda =0$, limit afterwards if desired.] Thus, we may indeed analyze (10) with the $1/2$ value (the transition from $\Delta$ to $D$ propagators is of course again smooth). Indeed, this requires no effort: Linearized gravity's consequences are well-enough known! They automatically imply the same miracle as in the Proca case: the radiation from the lower helicity modes is suppressed as 
$m$ vanishes, again upon proper normalization of the lower helicity actions by field redefinition.  More specifically, we first ``$3+1$" decompose the tensor $T_{\mu\nu}$ as 
$(T_{ij}, T_{0i},T_{00})$. The spatial part's equivalent of the vector decomposition (4) is [12]: 
\begin{equation}
T_{ij} = T_{ij}^{\hbox{\tiny{TT}}} + (T_{i,j} + T_{j,i}) + 1/2\, ( \delta_{ij}- \nabla^{-2} \dd_i \dd_j) T^T.
\end{equation}
The three sets of components $(T^{\hbox{\tiny{TT}}}, T^i, T^T)$ are again mutually orthogonal
decompositions of (tensorial) unity, numbering respectively $(2, 3,1)$ terms. The TT components are unconstrained by conservation, so subject to retarded interaction, in contrast to the seven components involved in conservation, $\dd^\mu\,  T_{\mu\nu} = 0$, namely $(T_i, T_{0\mu})$ which interact only instantaneously upon use of conservation. Significantly, the action (8) -- with the coefficient $1/2$ only --  contains no quadratic terms in the remaining, also unconstrained, scalar $T^T$, confirming absence of the corresponding field interaction in the massless limit; instead, $T^T$ just interacts instantaneously (according to Yukawa, rather than Coulomb, potentials of course) with $T_{00}$, as might be expected. These results are easy, if a bit tedious, to verify upon separating the usual linear action into its orthogonal parts, and repeated use of the various conservation constraints.

\section{Wave zone}

Before the above linearized conclusions can be applied to LIGO's observations, 
we must face the fact that emission took place in a (very) strong field regime, where linearization does not apply. While this is a simple problem in the Abelian regime, as in textbook Maxwell theory, setting up the concordance between strong interior emission and distant observation becomes a major problem in GR. Fortunately, it has been solved, in a necessarily lengthy technical paper [6], which dealt, in a (physically, at least) rigorous way, with the transition to asymptotic falloff of the strong field radiation excitations, unclouded by coordinate- and gravitational interaction-dependent issues. The upshot is that there is a well-defined spatial domain where physical modes behave sufficiently close to plane waves (with $1/r$ falloff, but derivatives also $\sim 1/r$) subject to linearized GR, with no details of the actual interior emission, though of course providing enough information to enable one to reconstruct the black hole merger as the natural candidate here. While even establishing the existence of a wave zone has never been attempted for the massive models, the conclusions may not differ substantially: the mass terms are relatively easy to tame and there are fewer ``coordinate wave" complications, given the loss of coordinate invariance here in favor of a fixed background; still, it must be proved. On the other hand, deciding if here too there is a corresponding strong field emission scenario is a truly daunting matter, without even hand-waving arguments: all that has been established is that massive black holes, whatever they may mean in these non-geometric models, have a spectrum similar to those of GR [15]. However, time-dependence, as in coalescence of two of them, is {\it terra incognita}; this is particularly true if one wishes to involve the Vainshtein mechanism [7], as emphasized in a relatively recent review [8].

\section{Conclusion}
A best-case view is that massive gravity may conceivably have a linear wave zone effectively decoupled from the strong interior, and subject to (linear) propagation, as is known to be true for GR. That is, for very small $m$, only quadrupole waves may emerge and be indistinguishable to $O(m)$ from GR's. If so, LIGO's observations would not distinguish directly between GR and its massive extensions. However, the wave zone amplitudes' information must also be traceable to an interior strong field emitting event. The violent black hole merger calculated to be responsible in GR has as yet no corresponding event in the massive models -- even what black holes and their mergers mean for $m\ne 0$ has never been studied: There might simply be no strong field event there that would generate the ring-down profile seen by LIGO. So it is possible that massive gravity has already been excluded by LIGO, absent an equally precise counterpart strong field event.   
 
\section*{Acknowledgements.}
This work was supported by grants NSF PHY-1266107 and DOE \# desc0011632. I thank J. Franklin for compositional dexterity.

%\begin{thebibliography}{99}
%\item  S. Deser, R. Jackiw, S. Templeton, ``Topologically Massive Gauge Theories  ", {\it Ann. Phys.} {\bf 140} (1982) 372.
\end{document}